\documentclass
[twocolumn,prl,superscriptaddress,tightenlines,lengthcheck]{revtex4}%
\usepackage{amsfonts}
\usepackage{amsmath}
\usepackage{amssymb}
\usepackage{graphicx}%
\providecommand{\U}[1]{\protect\rule{.1in}{.1in}}
\begin{document}
\preprint{cond-mat/0607244}
\title{Collapse of the critical state in superconducting niobium}
\author{Ruslan Prozorov}
\email{prozorov@ameslab.gov}
\affiliation{Ames Laboratory and Department of Physics \& Astronomy, Iowa State University,
Ames, Iowa 50011}
\author{Daniel V. Shantsev}
\affiliation{Department of Physics, University of Oslo, P. O. Box 1048 Blindern, 0316 Oslo, Norway}
\affiliation{A. F. Ioffe Physico-Technical Institute, Polytekhnicheskaya 26, St. Petersburg
194021, Russia}
\author{Roman G. Mints}
\affiliation{School of Physics and Astronomy, Raymond and Beverly Sackler Faculty of Exact
Sciences, Tel Aviv University, Tel Aviv 69978, Israel}

\begin{abstract}
Giant abrupt changes in the magnetic flux distribution in niobium foils were
studied by using magneto-optical visualization, thermal and magnetic
measurements. Uniform flux jumps and sometimes almost total catastrophic
collapse of the critical state are reported. Results are discussed in terms of
thermomagnetic instability mechanism with different development scenarios.

\end{abstract}
\date{10 July 2006}
\revised{20 September 2006}

\maketitle

Large jumps in magnetization, temperature, ultrasonic attenuation and
resistivity have been observed in many superconducting materials since early
$60$s \cite{mints,gurevich,altshuler,claiborne,goodman,levy}. Although no
definitive mechanism has been identified, it is believed that such changes are
caused by sudden redistribution of Abrikosov vortices triggered and supported
by thermomagnetic instabilities
\cite{mints,gurevich,altshuler,aranson,biehler,rakhmanov,denisov,shantsev,kuzovlev}%
. Direct observations of flux jumps using miniature Hall probes
\cite{james,altshuler2,esquinazi,chabanenko} and real-time magneto-optical
imaging \cite{leiderer,duran,altshuler,johansen,welling,wijngaarden} have
confirmed precipitous propagation of flux into the specimen. Although some
evidence of large - scale jumps exists \cite{goodman,james}, direct
observations of avalanche - like behavior in films have mostly found dendritic
and branched finger-like patterns (on a macroscopic scale of the whole sample)
or small jumps (few tens of micrometers) forming feather-shaped flux fronts
\cite{altshuler,altshuler2,duran,esquinazi,johansen,welling,wijngaarden}.

\begin{figure}
[b]
\begin{center}
\includegraphics[
height=6.9523cm,
width=9.1028cm
]%
{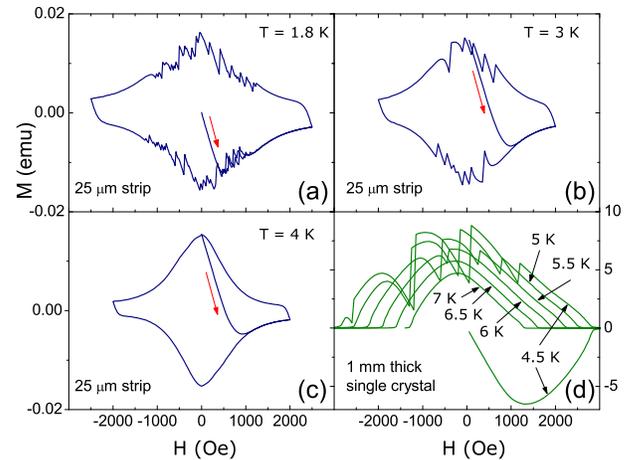}%
\caption{(color online) Magnetization loops measured in $25$ $\mu$m Nb foil at
different temperatures, (a) $1.8$ K, (b) $3$ K, (c) $4$ K and, for comparison,
several M(H)\ curves measured from $4$ kOe to $-4$ kOe in large Nb single
crystal. For (b) and (c) curves start from remanent magnetization, see text
for details.}%
\label{loops}%
\end{center}
\end{figure}

Dendritic avalanches are expected for thin films with high critical current
density \cite{aranson,kuzovlev,denisov}. The problem is that dendritic
avalanches are too small to result in the observed jumps of magnetization,
$M$. If $\delta M$ is a variation of magnetization during a flux jump, then
the dendritic avalanches contribute $\delta M\lesssim0.1M$ \cite{johansen}.
However, the observed changes are often much larger \cite{chabanenko,young}
(also see Fig. \ref{loops}). They must be associated with more dramatic
catastrophic changes of the critical state when a nonuniform (Bean's)
distribution of vortices suddenly collapses. In this paper we present direct
observations of the catastrophic collapse of the critical state in niobium
foils and provide evidence for thermomagnetic origin of this effect.%

%TCIMACRO{\FRAME{fbFU}{9.1028cm}{6.9523cm}{0pt}{\Qcb{(color online)
%Magnetization loops measured in $25$ $\mu$m Nb foil at different temperatures,
%(a) $1.8$ K, (b) $3$ K, (c) $4$ K and, for comparison, several M(H)\ curves
%measured from $4$ kOe to $-4$ kOe in large Nb single crystal. For (b) and (c)
%curves start from remanent magnetization, see text for details.}}{\Qlb{loops}%
%}{loops}{\special{ language "Scientific Word";  type "GRAPHIC";
%maintain-aspect-ratio TRUE;  display "ICON";  valid_file "F";
%width 9.1028cm;  height 6.9523cm;  depth 0pt;  original-width 4.8447in;
%original-height 3.6867in;  cropleft "0";  croptop "1";  cropright "1";
%cropbottom "0";  filename 'fig1.eps';file-properties "XNPEU";}}}%
%BeginExpansion
%EndExpansion

Pure Nb ($99.99$\%) strips of various thickness ($5$, $10$ and $25$ $\mu$m)
were obtained from \textit{Goodfellow}. Sample widths varied from $0.7$ to $4$
mm and the lengths varied from $2$ to $10$ mm. Magnetization loops, $M\left(
H\right)  $, were measured in a \textit{Quantum Design} MPMS. Resistivity,
specific heat and transport critical current were measured in a
\textit{Quantum Design} PPMS. We also measured sample temperature variation in
quasi-adiabatic conditions and found temperature jumps associated with the
flux jumps.%

%TCIMACRO{\FRAME{ftbFU}{9.1028cm}{3.8441cm}{0pt}{\Qcb{(color online)
%Penetration of magnetic field in $10$ $\mu$m thick Nb strip after zero-field
%cooling. (a) $160$ (b) $415$ (c) $480$ Oe. Penetration in (b) and (c) was
%avalanche-like at fields indicated. (Real-time video online \cite{webmovie}%
%).}}{\Qlb{zfc}}{zfc}{\special{ language "Scientific Word";  type "GRAPHIC";
%maintain-aspect-ratio TRUE;  display "ICON";  valid_file "F";
%width 9.1028cm;  height 3.8441cm;  depth 0pt;  original-width 3.5414in;
%original-height 1.478in;  cropleft "0";  croptop "1";  cropright "1";
%cropbottom "0";  filename 'fig2.eps';file-properties "XNPEU";}}}%
%BeginExpansion
\begin{figure}
[tb]
\begin{center}
\includegraphics[
height=3.8441cm,
width=9.1028cm
]%
{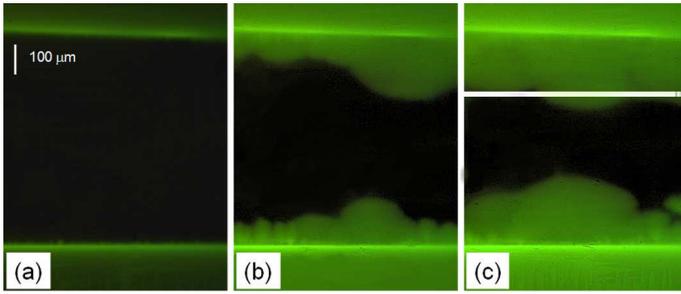}%
\caption{(color online) Penetration of magnetic field in $10$ $\mu$m thick Nb
strip after zero-field cooling. (a) $160$ (b) $415$ (c) $480$ Oe. Penetration
in (b) and (c) was avalanche-like at fields indicated. (Real-time video online
\cite{webmovie}).}%
\label{zfc}%
\end{center}
\end{figure}
%EndExpansion

Figure \ref{loops} shows $M\left(  H\right)  $ in a $25$ $\mu$m thick Nb foil
at different temperatures, (a) $1.8$ K, (b) $3$ K and (c) $4$ K. At the lowest
temperature, Fig. \ref{loops}(a), jumps in $M\left(  H\right)  $ are observed
both upon penetration after cooling in zero-field (ZFC) and upon flux exit. At
higher temperatures, jumps were only observed upon reduction of the magnetic
field, (b), and no jumps were observed above $4$ K, (c). In Fig.
\ref{loops}(b) and (c) curves are shown after the external field was reduced
from $5$ kOe to zero and an attempt was made to observe jumps upon increasing
the magnetic field of the same orientation. Unlike the case of the opposite
field, no jumps in $M\left(  H\right)  $ were observed. These results hint at
thermomagnetic instability mechanism \cite{mints,gurevich}. The comparison of
flux jumping instability scenarios in thin foils and bulk samples a large Nb
single crystal (5 mm diameter, 1 mm thickness) was measured at several
temperatures. Figure \ref{loops} (d) shows $M\left(  H\right)  $ at $T=1.8$ K
for both ascending and descending branches. For other temperatures only
descending branches are plotted. Giant magnetization jumps were observed upon
decrease of the magnetic field at temperatures up to $5.5$ K. Evidently,
variation of magnetization at a jump is much larger in a thick sample.%

%TCIMACRO{\FRAME{fbFU}{8.6042cm}{9.9881cm}{0pt}{\Qcb{Catastrophic collapse of
%the critical state upon entry of negative self-field at $T=3.9$ K. (a) $250$
%Oe (b) $200$ Oe (c) $100$ Oe (d) $50$ Oe. (Real-time video online
%\cite{webmovie}).}}{\Qlb{down}}{down}{\special{ language "Scientific Word";
%type "GRAPHIC";  maintain-aspect-ratio TRUE;  display "ICON";
%valid_file "F";  width 8.6042cm;  height 9.9881cm;  depth 0pt;
%original-width 3.5336in;  original-height 4.1502in;  cropleft "0";
%croptop "1";  cropright "1";  cropbottom "0";
%filename 'fig3.eps';file-properties "XNPEU";}}}%
%BeginExpansion
\begin{figure}
[b]
\begin{center}
\includegraphics[
height=9.9881cm,
width=8.6042cm
]%
{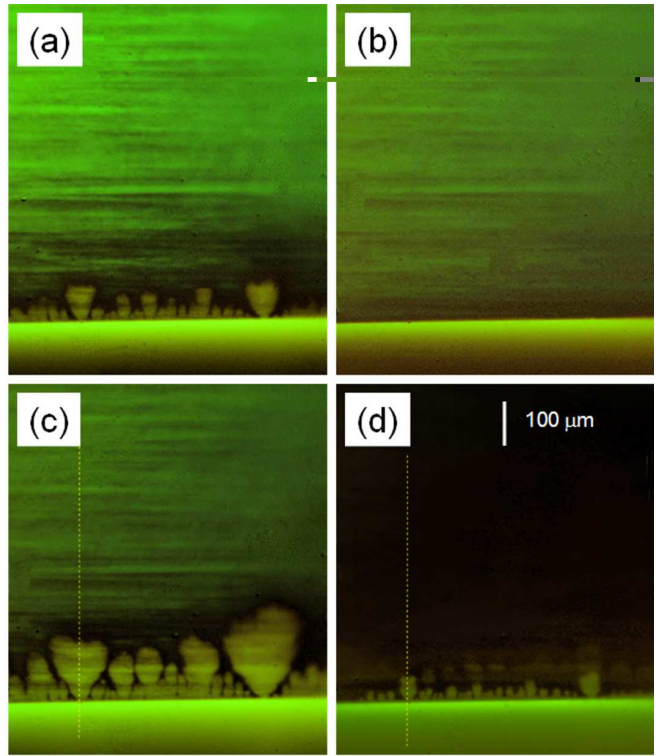}%
\caption{Catastrophic collapse of the critical state upon entry of negative
self-field at $T=3.9$ K. (a) $250$ Oe (b) $200$ Oe (c) $100$ Oe (d) $50$ Oe.
(Real-time video online \cite{webmovie}).}%
\label{down}%
\end{center}
\end{figure}
%EndExpansion

Visualization of the magnetic induction, $B\left(  r\right)  $, on the sample
surface was performed in a flow-type $^{4}$He cryostat with the sample in
vacuum attached to a copper heat exchanger (i.e., cooling conditions are
different from measurements in \textit{QD} MPMS, Fig.\ref{loops}).
Bismuth-doped iron-garnet heterostructure film with in-plane magnetization was
used as a magneto-optical indicator \cite{polyanskii,joss}. In all images,
intensity is proportional to the magnitude of $B\left(  r\right)  $. Moreover,
due to the dispersion of the Faraday rotation for different wavelengths, up
and down directions of the magnetic induction can be distinguished. In our
case, green corresponds to flux out of page and yellow corresponds to the
opposite orientation (antiflux). This information is useful for flux-antiflux
annihilation experiments described below.

In magneto-optical experiments, sudden collapses of magnetization were
observed only below $T\simeq4.5$ K for all samples. Figure \ref{zfc} shows
penetration of the flux into a $10$ $\mu$m Nb strip at increasing magnetic
fields, (a) $160$ Oe (b) $415$ Oe and (c) $480$ Oe. The large-scale abrupt
jumps in Fig. \ref{zfc} (b) and (c) occurred right before fields at which
images were acquired (about $405$ Oe and $470$ Oe, respectively). In our case
of 30 Hz data acquisition rate, the jumps seem instantaneous. Real-time
dynamics can be viewed online \cite{webmovie}.%

%TCIMACRO{\FRAME{fbFU}{8.6042cm}{7.1061cm}{0pt}{\Qcb{(color online) Profiles of
%the magnetic flux before and after collapse of the critical state upon entry
%of the negative flux generated by self-field. The profiles were measured along
%dashed lines in Fig.\ref{down} (c) and (d).}}{\Qlb{collapse}}{collapse}%
%{\special{ language "Scientific Word";  type "GRAPHIC";
%maintain-aspect-ratio TRUE;  display "ICON";  valid_file "F";
%width 8.6042cm;  height 7.1061cm;  depth 0pt;  original-width 4.2177in;
%original-height 3.4783in;  cropleft "0";  croptop "1";  cropright "1";
%cropbottom "0";  filename 'fig4.eps';file-properties "XNPEU";}}}%
%BeginExpansion
\begin{figure}
[b]
\begin{center}
\includegraphics[
height=7.1061cm,
width=8.6042cm
]%
{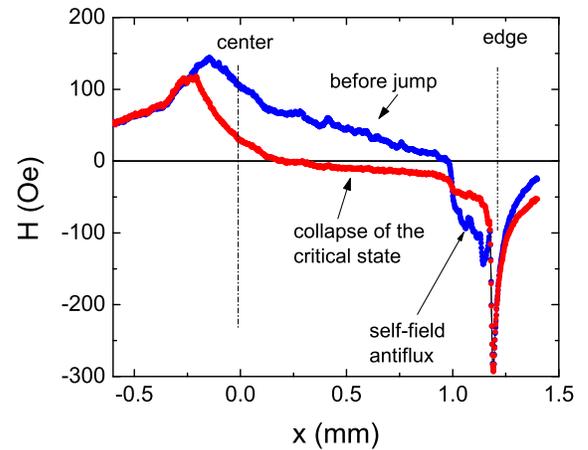}%
\caption{(color online) Profiles of the magnetic flux before and after
collapse of the critical state upon entry of the negative flux generated by
self-field. The profiles were measured along dashed lines in Fig.\ref{down}
(c) and (d).}%
\label{collapse}%
\end{center}
\end{figure}
%EndExpansion

The largest and most dramatic features occur upon decrease of a magnetic field
when self-field antiflux induced at the edges due to large demagnetization
factor starts entering the sample. The antiflux annihilates with the trapped
flux providing additional mechanism to trigger thermal instability. An
avalanche in progress is difficult to stop, because the whole sample is in the
critical state and an increase of temperature leads to reduction of the local
critical current density resulting in further flux redistribution. This
results in large uniform avalanches when the whole critical state literally
collapses. In Fig. \ref{down} two such collapses are shown, - panels (b)
(jumped after (a)), and (d) (jumped after (c)). Some intensity banding in Fig.
\ref{down} reflects the structure of the trapped flux in this sample.

\begin{figure}
[tb]
\begin{center}
\includegraphics[
trim=0.000000in 0.000000in 0.000708in 0.000000in,
height=5.4784cm,
width=9.1028cm
]%
{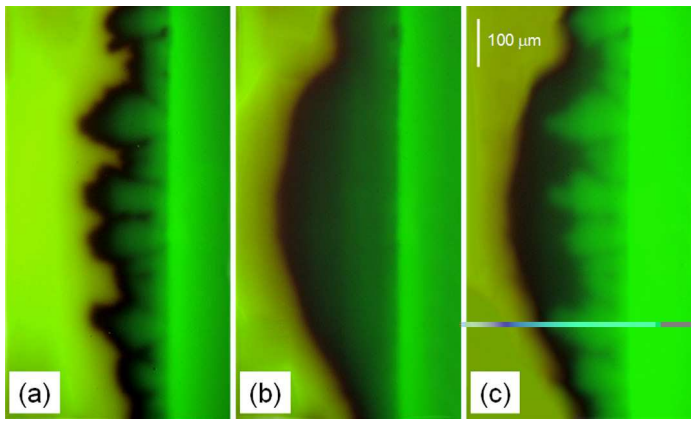}%
\caption{Entry of the antiflux (from right to left): (a) $-320$ Oe (b) $-470$
Oe. Note a huge uniform avalanche. (c) $-740$ Oe. Note secondary penetration
on top of the flux distribution flattened by the avalanche. It indicates that
$j_{c}$ returned to its large "cold" values. (Real-time video online
\cite{webmovie}).}%
\label{antiflux}%
\end{center}
\end{figure}

Flux profiles corresponding to the collapse of the critical state are shown in
Fig. \ref{collapse}. The profiles were measured along the dashed lines shown
in Fig. \ref{down} (c) and (d). Clearly, at some magnetic field, the slightest
variation results in a change of the entire distribution of vortices. The
observed giant jumps may span more than half the width of the simple, which,
to the best of our knowledge, has never been observed before. Moreover, the
collapse washes away both the interior critical state as well as any newly
penetrated antiflux at the edges. When the magnetic field of the opposite
polarity is applied, antiflux annihilates with the trapped flux similar to the
self-field antiflux discussed above. This results in large uniform avalanches
as shown in Fig. \ref{antiflux}. In general, the largest avalanches and
sometimes total collapse of the critical state occurs at small fields upon
entry of an antiflux. We think that this is because critical current density
is largest and most strongly field - dependent at low fields.%

%TCIMACRO{\FRAME{ftbFU}{9.1028cm}{5.4784cm}{0pt}{\Qcb{Entry of the antiflux
%(from right to left): (a) $-320$ Oe (b) $-470$ Oe. Note a huge uniform
%avalanche. (c) $-740$ Oe. Note secondary penetration on top of the flux
%distribution flattened by the avalanche. It indicates that $j_{c}$ returned to
%its large "cold" values. (Real-time video online \cite{webmovie}).}%
%}{\Qlb{antiflux}}{antiflux}{\special{ language "Scientific Word";
%type "GRAPHIC";  maintain-aspect-ratio TRUE;  display "ICON";
%valid_file "F";  width 9.1028cm;  height 5.4784cm;  depth 0pt;
%original-width 3.5414in;  original-height 2.1205in;  cropleft "0";
%croptop "1";  cropright "0.9998";  cropbottom "0";
%filename 'fig5.eps';file-properties "XNPEU";}}}%
%BeginExpansion
%EndExpansion

Although thermal instability seems to be the most natural scenario, direct
evidence for thermal effects is scarce. Temperature jumps were detected in
Nb-Zr alloys \cite{claiborne}. MgB$_{2}$ films coated with gold showed
suppression of the dendritic avalanches \cite{choi}. We report direct
measurements of the temperature variation associated with flux jumps in Nb
foils. In this experiment, the sample was suspended by four wires on a thin
sapphire stage normally used for specific heat measurements in a
\textit{Quantum Design} PPMS. The arrangement allows for direct reading of the
sample thermometer in almost adiabatic conditions.%

%TCIMACRO{\FRAME{ftbFU}{8.6042cm}{7.3741cm}{0pt}{\Qcb{(color online)
%Temperature variation upon decrease of magnetic field from $3$ kOe to $-3$ kOe
%in a $25$ $\mu$m thick Nb foil. The temperature variation was multiplied by a
%factor of $20$ for clear visual appearance.}}{\Qlb{avalanches}}{avalanches}%
%{\special{ language "Scientific Word";  type "GRAPHIC";
%maintain-aspect-ratio TRUE;  display "ICON";  valid_file "F";
%width 8.6042cm;  height 7.3741cm;  depth 0pt;  original-width 4.1779in;
%original-height 3.576in;  cropleft "0";  croptop "1";  cropright "1";
%cropbottom "0";  filename 'fig6.eps';file-properties "XNPEU";}}}%
%BeginExpansion
\begin{figure}
[tb]
\begin{center}
\includegraphics[
height=7.3741cm,
width=8.6042cm
]%
{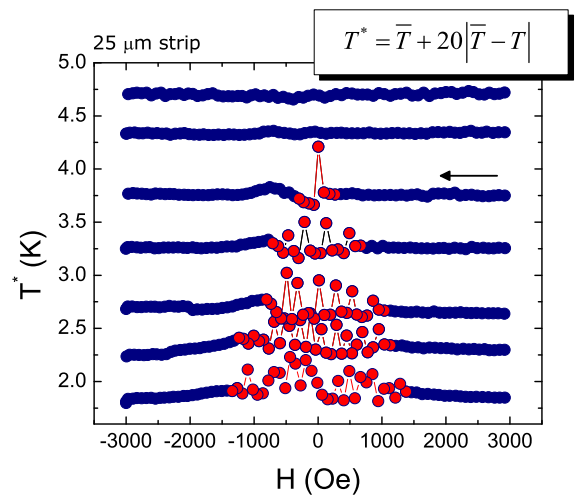}%
\caption{(color online) Temperature variation upon decrease of magnetic field
from $3$ kOe to $-3$ kOe in a $25$ $\mu$m thick Nb foil. The temperature
variation was multiplied by a factor of $20$ for clear visual appearance.}%
\label{avalanches}%
\end{center}
\end{figure}
%EndExpansion

Figure \ref{avalanches} shows field sweeps (all taken from $3$ kOe to - $3$
kOe). Temperature jumps were detected at the same range of temperatures and
fields as magnetization measured in the same sample. To emphasize the
variation, it was multiplied by a factor of $20$. Indeed, local temperature,
$T_{a}$, within the "volume" of the avalanche, $V_{a}\simeq5\times10^{-4}$
mm$^{3}$, reaches very large values of order of $T_{c}$ \cite{shantsev}. Then
the average temeperature measured by the calorimeter is given by $\delta
T=T_{a}V_{a}/V_{t}$, where total volume of the sample, $V_{t}\simeq
5\times10^{-2}$ mm$^{3}$. This gives $\delta T\simeq0.01T_{c}$ K, in agreement
with the direct measurements, Fig \ref{avalanches}.

In general, there are two channels to dissipate the energy of the critical
state -- dynamic and adiabatic \cite{mints,gurevich,aranson}. The dynamical
channel corresponds to the energy removal to a coolant and it is characterized
by the dimensionless parameter , $\alpha=\rho j_{c}^{2}d/h\left(  T_{c}%
-T_{0}\right)  $ \cite{mints,gurevich,aranson,kuzovlev,denisov}. Here $j_{c}$
is the critical current, $h\simeq5$ W/cm$^{2}$K, is the heat transfer
coefficient, $\rho\simeq0.5$ $\mu\Omega\cdot$cm is the normal state
resistivity, $d$ - thickness and $T_{0}$ - temperature of the coolant. The
adiabatic channel corresponds to the energy absorption by the material itself
which is accompanied by an increase of the sample temperature. The threshold
parameter for thin samples is given as $\beta=4dwj_{c}^{2}/\left[  c^{2}%
C_{p}^{v}\left(  T_{0}\right)  (T_{c}-T_{0})\right]  $ \cite{shantsev}, where
$w$ is the sample halfwidth. Using the results of direct measurements on our
foils, $j_{c}\left(  T\right)  =1.5\times10^{6}\left(  1-T/T_{c}\right)  $
A/cm$^{2}$ and $C_{p}\left(  T\right)  =0.25\left(  T/T_{c}\right)  ^{3}$
J/mol$\cdot$K, we obtain $\alpha\simeq8$ and $\beta\simeq15$. Large values of
$\alpha$ and $\beta$ indicate that both channels can take away only
insignificant fraction of the Joule heat, which makes flux jumps inevitable,
in full agreement with our experiments.

A flux jump in a thin sample develops into either dendritic or uniform pattern
depending on the ratio of thermal and magnetic diffusivities, $\tau
=\kappa/4\pi\rho C$, where $\kappa$ is the thermal conductivity.\cite{denisov}
Qualitatively, if $\tau$ is very small, the lateral heat diffusion is slow and
flux propagates along narrow channels forming a dendritic pattern. For large
$\tau$ the heat diffusion establishes essentially uniform temperature
distribution and a flux jump spreads out uniformly in all directions. The
threshold value is given (for $\alpha\gg1$) as $\tau_{c}\approx w/8dn\beta$
\cite{denisov}, where $n$ is the exponent in the current-voltage law,
$E\propto j^{n}$. For $n=30$ and $\kappa=10$~W/Km, we find $\tau\simeq1$,
while $\tau_{c}\simeq0.03$ indicating that the flux jumps should develop
uniformly -- as we indeed observe experimentally. To quantitatively validate
these estimates we considered development of flux jumps by numerically solving
the thermal diffusion and Maxwell equations, $C_{p}\ dT/dt=jE-h(T-T_{0})/d$
and $dE/dx=-dH/dt$. The sample was assumed to be a thin infinite strip, where
all the quantities depend only on the coordinate $x$ across the strip.

\begin{figure}
[t]
\begin{center}
\includegraphics[
height=7.5674cm,
width=9.1028cm
]%
{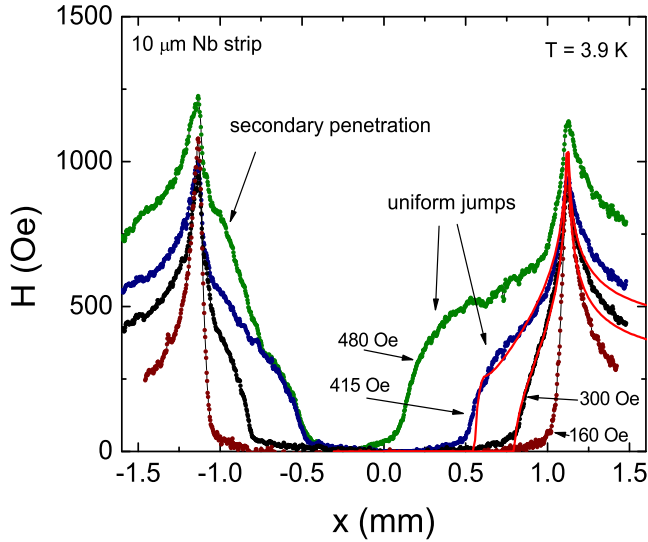}%
\caption{(color online) Profiles of magnetic induction in $2.88$ mm wide and
$10$ $\mu$m thick strip upon flux penetration after ZFC. Two inner profiles
correspond to large uniform jumps shown in Fig. \ref{zfc}.}%
\label{zfcprofiles}%
\end{center}
\end{figure}

Measured (symbols, corresponding to Fig. \ref{zfc}) and calculated profiles
(solid lines) of the magnetic induction are shown in Fig. \ref{zfcprofiles}.
If we omit the Joule term $jE$, the calculations produce the well-known
Bean-model profiles \cite{brandt} that perfectly fit the experimental profiles
before a jump, see the $300$~Oe curves. If we start from such a Bean profile
and, keeping applied field constant, introduce an infinitesimal perturbation
in $T$, a flux jump occurs. The curve for $415$ Oe shows the final flux
profile after such a jump. Apparently, the agreement with experiment is again
very good, which further supports the thermomagnetic mechanism of the observed instability.%

%TCIMACRO{\FRAME{ftFU}{9.1028cm}{7.5674cm}{0pt}{\Qcb{(color online) Profiles of
%magnetic induction in $2.88$ mm wide and $10$ $\mu$m thick strip upon flux
%penetration after ZFC. Two inner profiles correspond to large uniform jumps
%shown in Fig. \ref{zfc}.}}{\Qlb{zfcprofiles}}{zfcprofiles}%
%{\special{ language "Scientific Word";  type "GRAPHIC";
%maintain-aspect-ratio TRUE;  display "ICON";  valid_file "F";
%width 9.1028cm;  height 7.5674cm;  depth 0pt;  original-width 4.19in;
%original-height 3.4783in;  cropleft "0";  croptop "1";  cropright "1";
%cropbottom "0";  filename 'fig7.eps';file-properties "XNPEU";}}}%
%BeginExpansion
%EndExpansion

Furthermore, we tried to control the rate of heat conductance by introducing a
thin mylar film between the heat exchanger and the sample. Both,
feather-shaped protrusions and large uniform avalanches were observed in this
case. Moreover, uniform jumps appeared even in ZFC samples where they were
absent before. These experiments demonstrate that reducing the heat link to
the thermal bath makes the avalanches more likely to happen.

In conclusion, we observed large scale giant uniform vortex avalanches and
even collapse of the critical state in niobium foils. Direct measurements of
temperature variation as well as analysis of the magnetic and thermal energy
balance support thermomagnetic instability scenario. Comparable values of the
magnetic and thermal diffusivities ($\tau\sim1$) result in spatially uniform
giant flux jumps.

We thank Alex Gurevich, Igor Aranson and Yuri Galperin for useful discussions
and Thomas A. Girard for providing the samples and for discussions. Ames
Laboratory is operated for US DOE by the Iowa State University under Contract
No. W-7405-Eng-82. R. P. acknowledges support from NSF Grant DMR-05-53285 and
from the Alfred P. Sloan Research Foundation.

\end{document}